\documentclass[fleqn,twoside,twocolumn,nofootinbib,showkeys]{revtex4} 
\usepackage[nocpr,nopacs]{ujp}
\begin{document}
\title[Can Nuclear Matter Consist of $\alpha$-Particles?]
{CAN NUCLEAR\\ MATTER CONSIST OF $\boldsymbol\alpha$-PARTICLES?}%

\author{B.E.~Grinyuk}
\affiliation{Bogolyubov Institute for Theoretical Physics, Nat.  Acad. of Sci. of Ukraine}%
\address{14-B, Metrolohychna Str., Kyiv 03143, Ukraine}%
\email{bgrinyuk@bitp.kiev.ua}

\udk{539}  \razd{\seci}

\autorcol{B.E.\hspace*{0.7mm}Grinyuk}

\setcounter{page}{17}%

\begin{abstract}
A sufficient condition for the spatial collapse in an infinite
system of interacting Bose particles is obtained on the basis of
the variational principle with the use of trial functions with the
Jastrow pair correlation factors.\,\,The instability of a
hypothetical infinite system of $\alpha$-particles with respect to
the spatial collapse is shown under the assumption of the
Ali--Bodmer interaction potentials between such Bose
particles.\,\,Thus, it becomes clear why the hypothetical nuclear
matter is naturally treated with the use of at least the nucleon
degrees of freedom.
\end{abstract}

 \maketitle

\section{Introduction}

The $^{4}$He nucleus, or an $\alpha$-particle, is known to be a
strongly bound system of two protons and two neutrons (with the
binding energy of about 28.3~MeV) with zero spin and isospin.\,\,The
structure of an $\alpha$-particle is well-known (see, for example,
\cite{R1}).\,\,In\-side the nuclei, one can reveal the
$\alpha$-clusters.\,\,A number of light nuclei can be treated as a
definite number of interacting $\alpha$-particles (see \cite{R2} and
references ibid) or a system of $\alpha$-particles with additional
nucleons \cite{R3,R4,R5,R6}.\,\,There exist some approaches
\cite{R7} treating the nuclei (up to the heaviest ones) with even
and equal numbers of protons and neutrons as the systems of
\mbox{$\alpha$-particles.}

But what about the hypothetical nuclear matter: can one treat this
system as consisting of $\alpha$-particles?\,\,This question is
analyzed in the present paper, and it is shown that the system of
$\alpha$-particles is unstable with respect to the spatial collapse
if one ignores the Coulomb repulsion between the particles. To prove
this assertion from the very first principles, we consider a
Hamiltonian of $N$ interacting Bose particles with pairwise
interactions.\,\,We use the variational principle and derive a
sufficient condition for the spatial collapse in this system.\,\,The
obtained sufficient condition is a ge\-ne\-ralization of the one
obtained in \cite{R8}, and it is more refined.\,\,Using this
criterion, we show that an infinite system of $\alpha$-particles
with the known Ali--Bodmer interactions \cite{R9} between them is
unstable with respect to the spatial collapse.\,\,As it is clear
from the criterion formula given below, the same result is valid for
any other possible realistic versions of $\alpha$-$\alpha$
interactions of the same type with attraction at some distances of
about a few fermi and any non-singular repulsion at short
distances.\,\,It should be noted that, as usual in the nuclear
matter theory, we do not consider the Coulomb repulsion between
particles, otherwise the system will be destroyed by this long-range
repulsion and will not be bound at all.\,\,Just due to the Coulomb
repulsion, the nuclei heavier than those with $Z>\,\sim\!10^{2}$ are
not stable, nothing to say about
\mbox{$Z\rightarrow\infty$.}\looseness=1

In the next section, we obtain a sufficient condition for the
spatial collapse to take place in an infinite system of Bose
particles.\,\,We start with the Hamiltonian containing the
pairwise potentials, and then we use the Ritz variational
principle to derive this condition.\,\,As compared to the
criterion \cite{R8} obtained with the use of trivial trial
function in the form of the product of one-particle functions, we
now account for the Jastrow pair correlation factors and obtain a
more delicate criterion.

We stress that, as it was shown in \cite{R8}, the spatial collapse
of an infinite system of interacting Bose particles can not be
analyzed on the basis of the well-known Gross--Pitaevskii equation
\cite{R10,R11,R12} because the presence or absence of the spatial
collapse is not determined by the two-particle scattering length
\mbox{value.}\looseness=1

\section{Variational Estimation for~the~Energy~with~the~Use~of~Trial Functions~with~the~Jastrow~Factors}

We consider a system of $N$ identical interacting Bose particles of
mass $m$ with the Hamiltonian
\begin{equation}\label{E1}
\hat{H}=\sum_{k=1}^{N}\frac{\hat{\mathbf{p}}_{k}^{2}}{2m}+\sum_{n>k=1}^{N}V\left(\left|\bf{r}_{n}-\bf{r}_{k}\right|\right)\!,
\end{equation}
where the short-range pairwise potential depends on the distance
between particles.\,\,We assume that the profile of this potential
has the general form with some (possible) attraction of finite range
and some (possible) repulsion at short distances.

Now, we are going to find a sufficient condition for the spatial
collapse of the system under consideration to take place at
$N\rightarrow\infty$.\,\,In the previous paper \cite{R8}, we
proposed a simple sufficient condition of the spatial collapse in an
infinite system of interacting Bose particles to be $\int
V(r)d\mathbf{r}<0$.\,\,But, for a system of $\alpha$-particles, this
simple criterion appears to be too mild in order to decide
unambiguously whether one has the spatial collapse in an infinite
system of $\alpha$-particles (without the Coulomb repulsion).
Really, let us consider the $\alpha$-$\alpha$ interaction potential
in the form proposed by Ali and Bodmer \cite{R9}.\,\,Then, only for
two potentials denoted by $a_{0}$ and $b_{0},$ one obtains $\int
V\left(r\right)d\mathbf{r}<0$, i.e., the collapse takes place
without any doubts.\,\,For the rest versions of Ali--Bodmer
potentials, this sufficient condition is not fulfilled, and the
answer is undetermined.\,\,This only means that the used criterion
is to be improved.

In the previous paper \cite{R8}, we used the trial function in the
simplest form of a product of one-particle Gaussian
functions:\vspace*{-2mm}
\[
\Psi\left(\mathbf{r}_{1}, \mathbf{r}_{2},...,
\mathbf{r}_{N}\right)=\prod_{k=1}^{N}\exp\left(\!-\left(\mathbf{r}_{k}/R
\right)^{\!2}\right)\equiv
\]\vspace*{-7mm}
\begin{equation}\label{E2}
\equiv\exp\left(\!-\frac{1}{R^{2}}\sum_{k=1}^{N}\mathbf{r}_{k}^{2}\!\right)\!\!,
\end{equation}
where $R$ is a parameter of the order of the size of the
system.\,\,To improve the variational estimation, we now consider
the wave function with the Jastrow pair correlation factors taken
into account:
\begin{equation}\label{E3}
\Phi(\mathbf{r}_{1},\mathbf{r}_{2},...,
\mathbf{r}_{N})=\prod_{k>n=1}^{N}\!\!f(r_{kn})\Psi(\mathbf{r}_{1},\mathbf{r}_{2},...,
\mathbf{r}_{N}),
\end{equation}
where function $\Psi$ has the form (\ref{E2}), and the Jastrow
correlation factors $f(r_{kn})$ are chosen in the form
\begin{equation}\label{E4}
f(r_{kn})\equiv1-\exp(-r_{kn}^{2}/r_{0}^{2}),
\end{equation}
where $r_{nk}\equiv\left|\mathbf{r}_{n}-\mathbf{r}_{k}\right|$,
and $r_{0}$ is the radius of the order of the range of repulsion
in the pair potential $V(r)$ of interaction between
particles.\,\,Note that the choice of a specific form of the
correlation factors can change slightly the form of the criterion
to be obtained.\,\,But the chosen form (\ref{E4}) is sufficient
for our purpose to study the spatial collapse in an infinite
system of $\alpha$-particles with the Ali--Bodmer interaction
potentials.

To apply the Ritz variational principle (see, for instance,
\cite{R13}) for the ground-state energy of the system,\vspace*{-3mm}
\begin{equation}\label{E5}
E\,\leq\,\frac{\left\langle \Phi\,\right|\hat{H}\left|
\,\Phi\right\rangle}{\left\langle \Phi\, |\, \Phi\right\rangle},
\end{equation}
one has to calculate the matrix elements of the kinetic and
potential energies, as well as the matrix element $\left\langle
\Phi | \Phi\right\rangle$ in the denominator necessary to
normalize the trial function (\ref{E3}).\,\,To carry on
calculations in an explicit form, we consider, for simplicity, the
corresponding expressions in the limit $R\gg r_{0}$, where the
radius $R$ of the system is fixed, but much greater than the
radius $r_{0}$ of correlations.\,\,We also assume that $R$ is much
greater than the radius of short-range forces being of the same
order or compared with $r_{0}$.\,\,Thus, the limit $R\gg r_{0}$
means also that $R$ is much greater than the potential
radius.\,\,Then we consider the matrix element of the potential
energy in this limit:
\vspace*{-2mm}
\[
\left\langle \Phi \right|\!\! \sum_{n>k=1}^{N}\!\!V(r_{nk}) \left|
\Phi\right\rangle=\frac{N(N-1)}{2}\left\langle \Phi \right|
V(r_{12}) \left| \Phi\right\rangle\equiv
\]\vspace*{-4mm}
\[
\equiv\!\frac{N\!(N\!-\!1)}{2}\!\!\int \!\!
d\mathbf{r}_{1},...,d\mathbf{r}_{N}\Psi^{2}\!(\mathbf{r}_{1},...,
\mathbf{r}_{N})\, \times
\]\vspace*{-7mm}
\[
\times \,\,V\left(r_{12}\right)\!\!
\prod_{k>n=1}^{N}\!\!f^{2}(r_{kn}) \scriptsize{\begin
{array}{c}{\longrightarrow}\\{R\gg r_{0}}\end{array}}
\]\vspace*{-5mm}
\[
\scriptsize{\begin {array}{c}{\longrightarrow}\\{R\gg
r_{0}}\end{array}}\frac{N(N-1)}{2}C_{3,N}\!\!\int \!\!
d\mathbf{r}_{1}d\mathbf{r}_{2}f^{2}(r_{12})V(r_{12})\, \times
\]\vspace*{-7mm}
\begin{equation}\label{E6}
\times\,
\exp\left(\!-\frac{2}{R^{2}}(r_{1}^{2}+r_{2}^{2})\!\right)\!\!,
\end{equation}
where we used the notation
\[
C_{K,N}\equiv\int\!\!
d\mathbf{r}_{K},...,d\mathbf{r}_{N}\!\!\prod_{k>n=K}^{N}\!\!\!f^{2}\left(r_{kn}\right)\times
\]
\begin{equation}\label{E7}
\times\,
\exp\left(\!-\frac{2}{R^{2}}\sum_{k=K}^{N}\mathbf{r}_{k}^{2}\!\right)\!\!.
\end{equation}
In particular, the normalization matrix element $\left\langle \Phi |
\Phi\right\rangle\equiv C_{1,N}$ can be rewritten (at $R\gg r_{0}$)
as
\begin{equation}\label{E8}
C_{1,N}\scriptsize{\begin {array}{c}{\longrightarrow}\\{R\gg
r_{0}}\end{array}}\left(\!\frac{\pi}{2}\!\right)^{\!3}R^{6}\,C_{3,N}.
\end{equation}
If one use the new variables
$\mathbf{r}\equiv\mathbf{r}_{1}-\mathbf{r}_{2}$ and
$\boldsymbol{\rho}\equiv$
\mbox{$\equiv\frac{1}{2}(\mathbf{r}_{1}+\mathbf{r}_{2})$}, one
obtains, instead of (\ref{E6}),
\[
\left\langle \Phi \right|\!\! \sum_{n>k=1}^{N}\!\!\!V(r_{nk}) \left|
\Phi\right\rangle\scriptsize{\begin
{array}{c}{\longrightarrow}\\{R\gg r_{0}}\end{array}}
\]\vspace*{-7mm}
\begin{equation}\label{E9}
\scriptsize{\begin {array}{c}{\longrightarrow}\\{R\gg
r_{0}}\end{array}}\!
\frac{N\!(N\!-\!1)}{2}C_{3,N}\!\left(\!\frac{R}{2}\!\right)^{\!\!3}\!\!\pi^{3/2}\!\!\int
\!\!f^{2}(r)V(r)d\mathbf{r}
\end{equation}
in the limit $R\gg r_{0}$.\,\,Thus, the matrix element of the
potential energy $\left\langle \Phi \right| \hat{V} \left|
\Phi\right\rangle$ (\ref{E9}) divided by the normalization matrix
element (\ref{E8}) becomes as follows:
\begin{equation}\label{E10}
\frac{\left\langle \Phi \right|\!
\hat{V}\!\left|\Phi\right\rangle}{\left\langle \Phi |
\Phi\right\rangle}\!\scriptsize{\begin
{array}{c}{\longrightarrow}\\{R\gg r_{0}}\end{array}}\!
\!\frac{N(N-1)}{2}R^{-3}\pi^{-\frac{3}{2}}\!\!\int
\!\!f^{2}(r)V(r)d\mathbf{r}.\!\!\!\!\!\!\!
\end{equation}

To calculate the average of the kinetic energy, it may be suitable
to represent the matrix element of the kinetic energy operator in
the form
\[
\left\langle \Phi \right|\!
\sum_{k=1}^{N}\frac{\hat{\mathbf{p}}_{k}^{2}}{2m}\!\left|\Phi\right\rangle
\equiv-\frac{\hbar^{2}}{2m}\sum_{k=1}^{N}\left\langle \Phi
\right|\triangle_{k}\left|\Phi\right\rangle=
\]\vspace*{-6mm}
\begin{equation}\label{E11}
=-\frac{\hbar^{2}}{2m}N\left\langle \Phi
\right|\triangle_{1}\left|\Phi\right\rangle=\frac{\hbar^{2}}{2m}N\left\langle
\nabla_{1}\Phi |\nabla_{1}\Phi\right\rangle\!,
\end{equation}
where the gradient
\[
\nabla_{1}\Phi=\prod_{n>k=2}^{N}f(r_{nk})\exp\left(\!-\frac{1}{R^{2}}\sum_{s=2}^{N}r_{s}^{2}\!\right)\times
\]\vspace*{-6mm}
\begin{equation}\label{E12}
\times\nabla_{1}\left(\!\exp\left(\!-\frac{r_{1}^{2}}{R^{2}}\!\right)\prod_{j=2}^{N}f(r_{1j})\!\right)
\end{equation}
can be found explicitly. Substituting the result of differentiation
into (\ref{E11}) and integrating over $d\mathbf{r}_{1}$ and (for
convenience) over $d\mathbf{r}_{2}$, in the limit $R\gg r_{0}$, one
has
\[
\left\langle \Phi \right|\!
\sum_{k=1}^{N}\frac{\hat{\mathbf{p}}_{k}^{2}}{2m}\!\left|\Phi\right\rangle\!\!\scriptsize{\begin
{array}{c}{\longrightarrow}\\{R\!\!>\!\!>\!\!r_{0}}\end{array}}
\]
\begin{equation}\label{E13}
\!\scriptsize{\begin
{array}{c}{\longrightarrow}\\{R\!\!>\!\!>\!\!r_{0}}\end{array}}\frac{3}{8}\frac{\hbar^{2}}{2m}\pi^{3}\!R^{\,4}
C_{3,N}\!\cdot\!N\!\left(\!\!1\!+\!\frac{N\!-\!1}{2\sqrt{2}}\frac{r_{0}}{R}+\!o\!\left(\frac{r_{0}}{R}\right)\!\!\right)\!\!.
\end{equation}
Dividing this expression by the normalization matrix element in the
form (\ref{E8}) and adding (\ref{E10}), one ultimately has the
variational estimation for the energy of the ground state in the
limit $R\gg r_{0}$:
\begin{equation}\label{E14}
\frac{E}{N}\leq\frac{3}{2}\frac{\hbar^{2}}{mR^{2}}+\frac{N-1}{2R^{3}}\!\!\left(\!\frac{3}{2}\frac{\hbar^{2}}{m}
\frac{r_{0}}{\sqrt{2}}+\!\frac{1}{\pi^{\frac{3}{2}}}\!\!\int\!\!f^{2}\!(r)V\!\left(r\right)\!d\mathbf{r}\!\right)\!\!.
\end{equation}
It is obvious from the obtained expression that, under the
condition
\begin{equation}\label{E15}
A\equiv\frac{3}{2}\frac{\hbar^{2}}{m}
\frac{r_{0}}{\sqrt{2}}+\!\frac{1}{\pi^{\frac{3}{2}}}\!\!\int\!\!f^{2}\!\left(r\right)V\!\left(r\right)\!d\mathbf{r}<0,
\end{equation}
one has the spatial collapse in the system of interacting Bose
particles at $N\rightarrow\infty$.\,\,Real\-ly, in this case, the
energy (per one particle!) goes to minus infinity, as it is clear
from (\ref{E14}).\,\,At the same time, at a fixed parameter $R$, the
system of particles has a finite volume $\sim$$R^{3}$, but the
number of particles tends to infinity resulting in an infinite
density of particles.\,\,Note that the obtained sufficient condition
(\ref{E15}) generalizes our more simple criterion \cite{R8}
$\int\!V(r)\,d\mathbf{r}<0$, which follows from (\ref{E15}) at
$r_{0}\rightarrow 0$.

In the next section, we use the obtained criterion of the spatial
collapse in a Bose system to analyze whether a hypothetical system
of $\alpha$-particles can form a nuclear matter.\,\,There, the
negative answer will be obtained, since an infinite system of
$\alpha$-particles with typical $\alpha$-$\alpha$ interaction
potentials (without regard for the Coulomb repulsion) obeys
condition (\ref{E15}) of spatial collapse.

\section{The Spatial Collapse\\ of a Hypothetical Infinite System\\ of \boldmath$\alpha$-Particles without the Coulomb Repulsion}

Now, consider typical interaction potentials between
$\alpha$-particles in the form \cite{R9}
\begin{equation}\label{E16}
V\left(r\right)=V_{r}\exp\left(-\mu_{r}^{2}r^{2}\right)-V_{a}\exp\left(-\mu_{a}^{2}r^{2}\right)
\end{equation}

\begin{figure}
\vskip1mm
\includegraphics [width=\column] {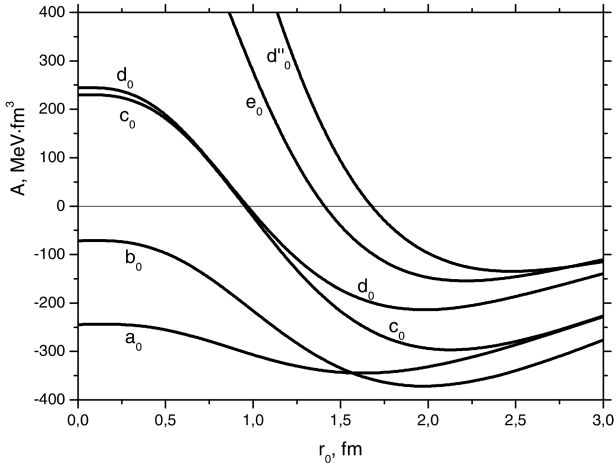}
\vskip-3mm\caption{Dependences of the left hand side $A$ of the
criterion (\ref{E15}) on the radius of correlations $r_{_0}$ for
different Ali--Bodmer $\alpha$-$\alpha$ potentials \cite{R9} (see
also the Table)}
\end{figure}

\begin{table}[b]
\noindent\caption{Parameters of some
$\boldsymbol\alpha$-$\boldsymbol\alpha$ potentials\\ from
\cite{R9}. The first column contains a notation \cite{R9}\\
of the corresponding version of a potential}\vskip3mm\tabcolsep4.7pt
\noindent{\footnotesize\begin{tabular} {|c|c|c|c|c|} \hline
\multicolumn{1}{|c}{\rule{0pt}{5mm}Potential}&
\multicolumn{1}{|c}{$\mu_{a}$ (fm$^{-1}$)} &
\multicolumn{1}{|c}{$V_{a}$ (MeV)} & \multicolumn{1}{|c}{$\mu_{r}$
(fm$^{-1}$)}& \multicolumn{1}{|c|}{$V_{r}$ (MeV)}\\[2mm] \hline
\rule{0pt}{5mm}$a_{0}$ & 0.35~\, & ~\,30 & 0.65 & ~\,125\\ $b_{0}$
& 0.42~\, & 150 & 0.55 & ~\,325\\ $c_{0}$ & 0.45~\, & 190 & 0.6~\,
& ~\,500\\
 $d_{0}$ & 0.475 & 130 & 0.7~\,  & ~\,500\\
$d\,''_{0}$ & 0.475 & 130 & 0.8~\,  & 1300\\ $e_{0}$ & 0.5~\,~\, &
150 & 0.8~\,  & 1050\\[2mm] \hline
\end{tabular}}
\label{Tab:PotenParam}%
\end{table}%

\noindent with a few sets of parameters for attraction and repulsion
given in the Table.\,\,It is worth to note that the
$\alpha$-$\alpha$ interaction \cite{R9} depends on the the angular
momentum.\,\,But, in the states with $l\neq 0$, the interaction
potentials are more attractive than in the state with $l=0$.\,\,We
simplify the problem and assume the potentials to be the same in all
the states.\,\,In this case, the interaction generally becomes a
little bit less attractive.\,\,If we shall demonstrate that, even
with such a simplified interaction, the hypothetical
$\alpha$-particle matter does collapse, then this effect should be
observed for original versions of $\alpha$-$\alpha$ interaction
\cite{R9} even more so.

Criterion (\ref{E15}) contains the radius of correlations $r_{0}$
which can be chosen in such a way that to make the contribution of
repulsion of the $\alpha$-$\alpha$ potential into the integral in
(\ref{E15}) sufficiently small.\,\,As a result, the integral of the
potential with the correlation factor squared becomes
negative.\,\,Due to the rather small contribution of the first term
originating from the kinetic energy matrix element, the correlation
factors at a definite $r_{0}$ turn the left-hand side of (\ref{E15})
to nega\-tive values at almost any short-range repulsion (except the
singular repulsion, in particular like ``hard core'').

In Figure, we show how the left-hand side $A$ of expression
(\ref{E15}) depends on $r_{0}$.\,\,It is seen that, for potentials
$a_{0}$ and $b_{0}$ from the Table, the criterion (\ref{E15}) is
fulfilled already at $r_{0}=0$.\,\,But, for the rest Ali--Bodmer
potentials, the inequality (\ref{E15}) is valid at nonzero $r_{0}$.
We do not depict the dependences at $r_{0}\rightarrow\infty$ for two
reasons.\,\,First, the term originating from the kinetic energy in
(\ref{E15}) is proportional to $r_{0}$, while the potential energy,
due to the increase of the radius of correlations, vanishes at
$r_{0}\rightarrow\infty$, and thus $A$ becomes positive at a
definite $r_{0}$ (depending on the version of
potential).\,\,Se\-cond, obtaining expression (\ref{E15}), we
assumed that $r_{0}\ll R$, where $R$ is fixed (although may be
rather large as compared to the radius of forces).\,\,Therefore, it
is not correct to tend the $r_{0}$ to infinity in expression
(\ref{E15}).

As it is clear, all the other possible versions of ``realistic''
local $\alpha$-$\alpha$ potentials should also give $A<0$ at a
definite radius of correlations $r_{0}$.\,\,Thus the hypothetical
system of $\alpha$-particles is unstable with respect to the spatial
collapse.\vspace*{-2mm}

\section{Conclusions}

To summarize, we note that a variational estimation with the account
for Jastrow pair correlation factors enabled us to demonstrate that
an infinite system of $\alpha$-particles can not form a hypothetical
nuclear matter due to its instability with respect to the spatial
collapse (if the Hamiltonian contains the Ali--Bodmer or similar
$\alpha$-$\alpha$ short-range interaction potentials and does not
contain the Coulomb long-range repulsion).\,\,We stress once more
that, as shown in \cite{R8}, the sign and value of the scattering
length of the pair potential can not be used as a criterion of the
effect of spatial collapse in Bose systems of interacting
\mbox{particles.}

The obtained criterion of the spatial collapse of an
infinite system of Bose particles can be used for studying a
possible spatial collapse in other physical systems including
imperfect Bose gases.

\vskip2mm
{\it This work is partially supported by the National Academy of
Sciences of Ukraine, Project No.\,\,0121U114399.}

\rezume {Б.Є.\,Гринюк}{ЧИ МОЖЕ ЯДЕРНА\\ МАТЕРІЯ СКЛАДАТИСЯ З
$\alpha$-ЧАСТИНОК?}{На основі варіаційного принципу з використанням
пробних функцій з парними кореляційними факторами Ястрова отримано
достатню умову просторового колапсу в нескінченній системі
взаємодіючих бо\-зе-час\-ти\-нок. Показано, що гіпотетична
нескінченна система $\alpha$-час\-ти\-нок нестабільна по відношенню
до просторового колапсу у припущенні потенціалів взаємодії
Алі--Бод\-ме\-ра між такими бо\-зе-час\-тин\-ками. В результаті стає
зрозумілим, чому гіпотетичну ядерну матерію природно розглядати
принаймні з використанням нуклонних ступенів вільності.}
{\textit{К\,л\,ю\,ч\,о\,в\,і\, с\,л\,о\,в\,а:}\, бозе-система з
$\alpha$-частинок, просторовий колапс, ядерна матерія.}

\end{document}